\documentclass[twocolumn,showpacs,floatfix,aps,pra,superscriptaddress]{revtex4}
\usepackage{graphicx}
\usepackage{dcolumn}
\usepackage{bm}

\begin{document}

\title{Effects of the interplay between interaction 
and disorder in bipartite entanglement}

\author{L. F. Santos}
\email{santos@pa.msu.edu} 
\affiliation{Department of Physics and Astronomy, Michigan State
University, East Lansing, MI 48824, USA}
\author{G. Rigolin}
\email{rigolin@ifi.unicamp.br}
\affiliation{Departamento de Raios C\'osmicos e Cronologia, Instituto de 
F\'{\i}sica Gleb Wataghin, Universidade Estadual de Campinas, C.P. 6165, 
cep 13084-971, Campinas, S\~ao Paulo, Brazil}

\begin{abstract}
We use a disordered anti-ferromagnetic spin-1/2 chain with 
anisotropic exchange coupling to model an array of interacting qubits.
All qubits have the same level spacing, except two, which are called
the defects of the chain. The level spacings of the defects 
are equal and much larger than all the others.
We investigate how the entanglement between the two defects
depends on the anisotropy
of the system. When the anisotropy coupling is much larger than the
energy difference between a defect and an ordinary qubit, 
the two defects become strongly entangled. Small anisotropies, on the contrary,
may decrease the entanglement, which is, in this case, also much affected 
by the number of excitations.
The analysis is made for nearest neighbor and next-nearest neighbor defects.
The decrease in the entanglement for nearest neighbor defects is not very
significant, especially in large chains.

\end{abstract}


\maketitle

\section{Introduction}

Entanglement is one of the most striking properties of
quantum mechanics. It describes a correlation between quantum mechanical
systems that does not occur in classical mechanics, namely that
a pure state of a composite quantum system cannot be written as a
product of the states of its constituents \cite{Schrodinger}. 
It has long been a subject of interest in
foundations of quantum mechanics, but recently
it has acquired a fundamental role in quantum computation and quantum 
information \cite{chuang}.

Several attempts to quantify entanglement have been developed \cite{measures},
but this is an area of research still in progress.
In the case of a pure state of a bipartite system, a
good and widely accepted measure of entanglement is the von Neumann
entropy of its reduced density matrix. This entropy 
can be associated with a quantity
called concurrence \cite{wootters}, which we adopt here as our
measure of entanglement. The concurrence varies from 0 to 1, the larger 
it is the more entangled is the state. 

Spin chains are ideal systems for the study of entanglement.
They are naturally used to model quantum computers (QC's): the two
states of a spin-1/2 particle correspond to the two levels of a qubit
and the exchange interaction corresponds to the qubit-qubit interaction.
A particle with spin up corresponds to an excitation  
or an excited qubit. We consider a spin chain described by the $XXZ$ model. This is the model used in quantum computers based on electrons on helium \cite{dyk}.
The Ising part of the interaction is proportional to the anisotropy
coupling, as shown in the next section. It only plays a role when two
or more excitations are present in the chain.
The XY part of the interaction is responsible for hopping the excitations and
can be used to create entanglement
between two or more qubits \cite{chines,pra}. However it can 
also quickly destroy bipartite entanglement, because the excitations
soon spread all over the chain.

A way to avoid the spreading (or delocalization) of the excitations is
by tuning the qubits away from resonance \cite{perpetual}.
Such control of the qubit level spacings 
enables us to entangle just some specific qubits. 
This is the strategy we use in this paper.
To have two chosen qubits maximally entangled, we tune them in resonance.
Their level spacings are different from all other qubits, they are the 
defects of the chain.

When there is only one excitation in the chain, the entanglement
between the two selected sites becomes trivial.
By assuming that the energy difference between the defects and the 
ordinary qubits is much larger than the strength of
the $XY$ interaction, the Hamiltonian of the system gives
two maximally entangled states.
They correspond to superpositions of the basis states where the
excitation occupies one of the defects \cite{pra}.
However, when more than one excitation is present, the 
entanglement between the defects can be significantly affected by
the Ising interaction. 
The main purpose of the present work is
to analyze which is, in this case, the eigenstate with maximum concurrence 
and how the value of the maximum concurrence depends on 
the anisotropy.

We find numerically that for large values of the a\-ni\-so\-tro\-py coupling, 
nearest neighbor defects are strongly entangled and the
value of the concurrence is almost negligibly affected by the number 
of excitations in the system. In the case of separated defects, this 
observation is valid only for long chains. In the opposite situation 
of small anisotropies, the concurrence between the defects is 
strongly dependent on the number of excitations and it can become quite
small. Partial analytical justifications for these observations are
provided.

The paper is organized as follows. In Sec. II we describe the model
and the adopted measure for bipartite entanglement. In Sec. III
we analyze the maximum concurrence between nearest neighbor and next nearest 
neighbor defects for various values of the anisotropy coupling.
A discussion on how to prepare 
maximally entangled states and how the anisotropy may help
is also presented in this section.  

\section{The Model and The Measurement of Entanglement}

We consider a spin chain with nearest neighbor interaction. 
The Hamiltonian describing the system is given by the $XXZ$ model

\begin{eqnarray}
\label{XXZ}
&&H = \sum _{n=1} ^{L} \frac{\varepsilon _n}{2} \sigma _{n}^{z} + 
\frac{J}{4} \sum _{n=1} ^{L} \left[ \Delta  
\sigma _{n}^{z} \sigma _{n+1}^{z} + \frac{1}{2} H_{\rm{hop}} 
\right], \\ 
&&H_{\rm{hop}} = \left(\sigma _{n}^{+} 
\sigma _{n+1}^{-} + \sigma _{n}^{-} \sigma _{n+1}^{+}  \right) 
\nonumber,
\end{eqnarray}
where $\hbar =1$ and $\sigma ^{z,+,-}$ are Pauli matrices.
There are $L$ sites and we deal with a periodic (or closed) chain, that is, 
sites $n+L$ and $n$ are the same. Each site $n$ is subjected to a magnetic 
field in the $z$ direction, giving the energy splitting $\varepsilon _n$. In 
this description, the excitation energy of a qubit is the Zeeman 
energy of a spin. In this disordered chain,
not all the qubits have the same level spacing
$\varepsilon $. The energy of two qubits, called defects, differs from those
of other qubits by $d$.
The parameter $J$ is the hopping integral and $\Delta $ is 
a dimensionless parameter related to the anisotropy 
coupling. The diagonal term $\sigma _{n}^{z} \sigma _{n+1}^{z}$ 
gives the Ising interaction and the non-diagonal term $H_{\rm{hop}}$ 
is responsible for propagating the excitations. We set 
$d,J$ and $\Delta>0$.

The anisotropy in the Hamiltonian (\ref{XXZ}) is different
from the one considered in some previous models \cite{osterloch,osenda,huang}. 
The hopping part of the Hamiltonian can be equivalently written as 
$H_{\rm hop}\propto 
J_x \sigma_n^x \sigma_{n+1}^x + J_y \sigma_n^y \sigma_{n+1}^y$. 
Here $J_x=J_y=J$, but in the models
cited above, the degree of anisotropy comes from the difference
between $J_x$ and $J_y$. The anisotropy in our case originates from 
the extra Ising interaction. In terms of entanglement, few studies
have been developed with this model \cite{rigolinIJQI,russian}. 
Moreover, our goal is not simply
to analyze how entanglement may
depend on the described anisotropy, but we aim to analyze how
it is affected by the interplay between this interaction and disorder. 

The disorder of the system we consider is characterized by the presence
of the two defects. In principle, total control of the qubit level
spacings is available, which allows the creation of defects.

The Ising part of the $XXZ$ model combined with defects has been used before 
to create maximally entangled states \cite{pra}. The study of entanglement
with impurities, but without any $\sigma_n^z \sigma_{n+1}^z$ 
interaction, has also been done in 
\cite{osenda,huang}. We emphasize the importance of studying
the effects of this extra 
interaction in disordered systems, which are far from trivial. The Ising 
interaction is only
relevant when at least two excitations are present, being therefore
associated with many-body problems. It is actually at the heart of one
of the most challenging problems in condensed matter physics, namely
the difficulty in localizing many-particle states \cite{perpetual}.

In what follows, we count energy off the ground state energy
$E_{0} = -(L \varepsilon + 2 d)/2 + L J\Delta/4$, i.e., we 
replace in Eq.~(\ref{XXZ}) $H\rightarrow H-E_0 $. 

To address the different states of the system we use a notation that is
common in the study of spin chains with the Bethe ansatz \cite{bethe}.
The state corresponding to one single excitation on site $n$,
that is  $|\downarrow _{1} \downarrow _{2}...\downarrow _{n-1} 
\uparrow _{n} \downarrow _{n+1}... \downarrow _{L}\rangle $,
or equivalently $|0_{1} 0_{2}... 0_{n-1}1_{n}0_{n+1}... 0_{L}\rangle $, is
simply written as $\phi (n)$. 
The state of two excitations, one on 
site $n$ and the other one on site $m$, is $\phi (n,m)$, 
which is a simplified notation for  
$|\downarrow _{1} \downarrow _{2}...\uparrow _{n} \downarrow _{n+1}...
\uparrow _{m} ... \downarrow _{L}\rangle $, or equivalently
$|0_{1} 0_{2} ...1_{n} 0_{n+1}... 1_{m} ... 0_{L} \rangle $.  
Basis states where each excitation is confined to a
single site, such as $\phi(n,m)$, are called in quantum 
computing quantum registers. 
These are the states where 
the measurements are performed.

Since we want to study the effects of the Ising interaction, 
several excitations have to be considered, which 
limits the numerical analysis to small chains.
In the model described by Eq.~(\ref{XXZ}), the $z$ component of the total
spin $\sum_{n=1}^{L} S_{n}^{z}$ is conserved, so states with different
number of excitations are not coupled and the Hamiltonian is made of 
uncoupled blocks. However the blocks can still be very large. For a chain with 
$L$ sites, the block corresponding to $N$ excitations
has dimension $L!/[N!(L-N)!]$. Its diagonalization leads to eigenstates which
correspond to linear superpositions
of quantum registers $\phi$ with $N$ excitations. 
Each state $k $ is written as

\begin{equation}
|\psi ^{(k )}_{N,L}\rangle = 
\sum_{n<m<... <N=1}^L a^{(k)}(n,m,..., N) |\phi (n,m,..., N)\rangle .
\end{equation}

In order to study quantitatively the entanglement between two qubits 
we calculate their entanglement of formation $E_{F}$. Given the 
density matrix $\rho$ that describes our pair of qubits, $E_{F}$ is the 
average entanglement of the pure states of the decomposition of $\rho$, 
minimized over all possible decompositions:
\begin{equation}
E_{F}(\rho) = \text{min}\sum_{i}p_{i}E(\psi_{i}),
\end{equation}
where $\sum_{i}p_{i} = 1$, $0 < p_{i} \leq 1$, and 
$\rho = \sum_{i}p_{i}\left| \psi_{i}\right>\left< \psi_{i}\right|$. 
Here $E(\psi)$ is the von Neumann entropy of either of the two qubits 
\cite{bennett}. Wootters \textit{et al} \cite{wootters} have shown that, 
for a pair of qubits, $E_{F}$ is a monotonically increasing function of 
the concurrence, which one can prove to be an entanglement monotone. Since 
the concurrence is mathematically simpler to deal with than $E_{F}$, we 
adopt it here to measure the
entanglement between two qubits. It is given by 
\cite{wootters}:
\begin{equation}
C = \text{max} 
\{ \lambda_{1} - \lambda_{2} - \lambda_{3} - \lambda_{4}, 0 \},
\end{equation}
where $\lambda_{1},  \lambda_{2}, \lambda_{3}$, and $\lambda_{4}$
are the square roots of the eigenvalues, in decreasing order, of the matrix 
$R =\rho\tilde{\rho}$. The matrix $\tilde{\rho}$ is the time reversed 
matrix 
\begin{equation}
\tilde{\rho} = \left(\sigma_{y} \otimes \sigma_{y}\right) \rho^{*} 
\left(\sigma_{y} \otimes \sigma_{y}\right).
\end{equation} 
The symbol $\rho^{*}$ means complex conjugation of the matrix $\rho$ in the 
basis $\left\{ \left| 11 \right>, \left| 10 \right>, \left| 01 \right>, 
\left| 00 \right> \right\}$. 

Maximum entanglement corresponds to $C=1$ and no entanglement gives
$C=0$. To compute the concurrence of two qubits in a chain with several
sites, we
trace over the qubits we are not interested in and study the reduced density 
matrix of the two chosen ones.

\section{Numerical Results and Analytical Analysis}

We study the entanglement between the two defects of the chain.
The difference in energy between them and the other qubits,
$d$, is assumed much larger than the hopping integral, $d>>J$. 
This guarantees that
an excitation placed on one defect can only hop between the 
two defect sites. As a consequence, among all the eigenstates of the system,
the ones with a single excitation shared between the
defects have the largest concurrences.

In the first subsection below, we analyze the 
case where the defects correspond to two nearest neighbor
qubits and in the second subsection they are two next nearest neighbors. 
Since the
chain has periodic boundary conditions, any pair of qubits $(n_0, m_0)$
translated through the chain 
is equivalent, so we choose, for the numerical calculations, 
the pair 1 and 2 as nearest neighbor defects and
the pair 1 and 3 as next-nearest neighbor defects.

We study how the maximum concurrence between the defects depend
on the anisotropy and which is the corresponding eigenstate.

\subsection{Nearest neighbor defects}

\begin{figure}[!ht]
\includegraphics[width=2.5in]{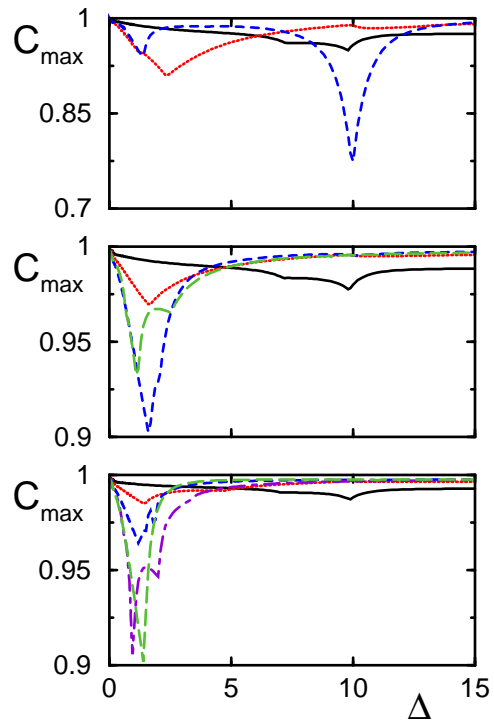}
\caption{(color online)
Maximum concurrence vs. anisotropy coupling
in closed Heisenberg spin-1/2 chains
with defects located on sites 1 and 2. The energy difference between
the defects and the other qubits is $d=10J$. 
The panels correspond to data for 
chains of different lengths: $L=8$ (top), $L=10$ (middle) and 
$L=12$ (bottom). The maximum number of excitations considered for each
chain is $L/2$. The curves are chosen as follows: 
solid (black) curves for 2 excitations,
dotted (red) for 3 excitations, dashed (blue) for 4 excitations,
long dashed (green) for 5 excitations, and 
dot-dashed (violet) for 6 excitations.}
\label{fig1}
\end{figure}

The numerical results for the dependence of the maximum concurrence,
$C_{\max}$,
on the anisotropy of a periodic Heisenberg chain with two 
neighbor defects are
shown in Fig.~\ref{fig1}. When $\Delta =0$, all 
states with one excitation on the defects have $C_{\max}\simeq 1$.
For small $\Delta $'s, in general, $C_{\max}$ decreases with the 
number of excitations. 
The comparison of $C_{\max}$ for
chains of different sizes but with the same 
number of excitations indicates that, in most cases, smaller chains are 
more affected by the Ising interaction.
When there are just two excitations, the minimum value of 
$C_{\max}$ happens when $J\Delta = d$, while
in the case of more excitations, this occurs for smaller values of $\Delta $.
Whenever $J\Delta \gg d$, the maximum concurrence 
stabilizes in a value close to 1 (the 
larger the chain the closer to 1 it will be). In this case, 
the dependence on the number of
excitations becomes little noticeable.
In what follows, we try to find justifications, some times analytically,
to these observations.

The eigenvalues and eigenstates of an anisotropic spin chain with no
defects can be analytically obtained with the Bethe ansatz \cite{bethe}.
In the presence of defects, where $d\sim J$, most commonly the chain 
becomes non integrable \cite{jpa_me,pra_escobar}. When $d\gg J$, as in
this paper, we are again capable of solving the eigenvalue problem in 
certain situations. 
This happens 
because, with such large $d$, the $XY$-type interaction between defect
and ordinary qubit is negligible, there is no hopping of excitations 
between them. It works as if we had cut the closed chain and created two
chains with free boundaries (two open chains):
a small one corresponding simply to the two defects and another
one with $L-2$ sites.

In the simple case of a single excitation in the chain, the spectrum is
divided into two well separated bands. 
One is made of $L-2$ states with energy in the range 
$\varepsilon_1 -J \leq E_1\leq \varepsilon_1 +J$,
where $\varepsilon _1 = \varepsilon -J\Delta $. They are states
with the excitation out of the defects. The other band is made of
just the following two linear superpositions 
$[\phi(n_0) \pm \phi(n_0+1)]/\sqrt{2}$, with energy 
$\varepsilon _1 + d \pm J/2 $, respectively. They correspond to 
Bell (or EPR) states and have concurrence equal to $1$.

In this paper we want to analyze how the Ising interaction
affects the entanglement and so we consider two or more excitations.

\subsubsection{Two excitations and $\Delta =0$}

In the case of two excitations, but with $\Delta =0$,
we should obtain results similar to ones for one excitation,
that is the states with maximum concurrence should give $C_{\max}=1$,
since no Ising interaction exists. 
The states of maximum concurrence between the defect sites have
one excitation shared between them. They are superpositions of
the quantum registers with energy $\varepsilon_2 + d$, where 
$\varepsilon _2 = 2 \varepsilon -2 J\Delta $ and form an energy band
that we will call defect band, or $d$-band for short. 
The transitions between these quantum registers 
are schematically shown below. Each 
quantum register (apart from the border ones)  is
coupled to three others: two corresponding to an
excitation hop to an ordinary qubit
and one associated to an excitation hop to the other defect. 

\footnotesize
\[
\begin{array}{cccccc}
n_0,n_0\!+\!2& \leftrightarrow & n_0,n_0\!+\!3& \leftrightarrow & 
... &n_0,n_0\!+\!L\!-1\\

\updownarrow&   & \updownarrow & & &\updownarrow \\

n_0\!+\!1,n_0\!+\!2& \leftrightarrow & n_0\!+\!1,n_0\!+\!3& \leftrightarrow & 
... &n_0\!+\!1,n_0\!+L\!\!-\!1

\end{array}
\]
\normalsize

The states of the $d$-band have the form

\begin{eqnarray}
|\psi^{(k_1,k_2)}_{2,L} \rangle &=& \sum_{n,m} c(n,m)
|\phi (n,m)\rangle 
\nonumber \\
&=& \sum_{m=n_0+2}^{n_0-1+L} c(n_0,m) |\phi (n_0,m) \rangle \nonumber \\
&+& 
\sum_{m=n_0+2}^{n_0-1+L} c(n_0+1,m) |\phi (n_0+1,m) \rangle .
\label{psi2}
\end{eqnarray}
[To avoid a heavy notation, the superscript 
$(k_1,k_2)$ was suppressed from the right hand side.].

We use the Bethe ansatz method to find the coefficients $c(n,m)$ 
for these $2(L-2)$ states.
The Schr\"odinger equation for $c(n,m)$ is

\begin{eqnarray}
\label{e2}
E_2 c(n,m)& = & (\varepsilon_2 +d) c(n,m) + \frac{J}{2} [c(n-1,m)  \nonumber \\
& + & c(n+1,m) + c(n,m-1) + c(n,m+1)]. \nonumber \\
&&
\end{eqnarray} 

Since the studied chain became equivalent to two
open chains, it is natural to write

\begin{eqnarray}
&& c(n,m) = a(n) b(m) = \nonumber \\
&& [A_{1} \mathrm{e}^{\mathrm{i}\alpha n} +A_{2} \mathrm{e}^{-\mathrm{i}\alpha n}][B_{1} \mathrm{e}^{\mathrm{i}\beta m} +B_{2} \mathrm{e}^{-\mathrm{i}\beta m}].
\label{coef}
\end{eqnarray}

The energy $E_2$ as a function of $\alpha $ and $\beta $
is obtained from (\ref{e2}), (\ref{coef}) and has the form

\begin{eqnarray}
\label{energy}
E_2 &=& \varepsilon_2 + d +E_a + E_b, \\
E_a &=& J\cos\alpha, \nonumber \\
E_b &=& J\cos \beta. \nonumber
\end{eqnarray}
Using the boundary conditions for the small chain of
defects

\begin{equation}
a(n_0-1)=a(n_0+2)=0,
\label{bound_a}
\end{equation}
and the boundary conditions for the chain of ordinary qubits
\begin{equation}
b(n_0+1)=b(n_0+L)=0,
\label{bound_b}
\end{equation}
we obtain 
\[
c(n,m) = A \sin[\alpha (n - n_0 +1)] 
\sin [\beta (m - n_0 -1)], 
\] 
where $A$ is a normalization constant and $\alpha $ and $\beta $
are given by

\begin{eqnarray}
\label{alpha}
&&\alpha = \pi k_{1}/3, \quad  k_{1}=1,2; \\ 
&&\beta = \pi k_{2}/(L-1), \quad  k_{2}=1,2,...,L-2. 
\end{eqnarray}
Each one of the $2(L-2)$ states have then a ``Bell-type'' form 

\begin{eqnarray}
&&|\psi(n,m)^{(k_1,k_2)}\rangle = \sum_{m=n_0+2}^{n_0-1+L}
A \sin \left[ \frac{\pi k_2 (m - n_0 -1)}{(L-1)}\right] \nonumber \\
&&\times \frac{1}{\sqrt{2}}\left[ \phi(n_0,m)
+(-1)^{k_1 +1} \phi(n_0 +1,m) \right].
\label{bell}
\end{eqnarray}

To calculate the concurrence between the two defects, 
we obtain the reduced density matrix of the states above 
by tracing over the
ordinary qubits. Only 
the four elements in the middle are different from zero.
We have

\begin{equation}
  \rho =  
    \left(
	\begin{array}{cccc}
            0 & 0 & 0 & 0 \\
	    0 & 1/2 & (-1)^{k_1 +1}/2 & 0 \\
            0 & (-1)^{k_1 +1}/2 & 1/2 & 0 \\
	    0 & 0 & 0 & 0   	
	\end{array}
    \right).
\end{equation}
Therefore, when $\Delta =0$, all the states in the $d$-band 
have concurrence equal to $1$. These analytical results agree very well
with the numerical ones.

\subsubsection{Two excitations and $0<\Delta < d/J$}

The situation changes as we start increasing $\Delta $.
The scheme for the involved transitions from the previous sub-subsection
still applies, but
the quantum registers $\phi (n_0-1, n_0)$ and $\phi(n_0 +1, n_0+2)$
now have on-site energy equal to $\varepsilon_2 + d + J\Delta $, 
which is larger than the energy
of the other quantum registers considered.
As a consequence, the coefficients $c(n,m)$ are not simply written as
the product (\ref{coef}), which leads to the ``Bell-type'' states given
by Eq.(\ref{bell}). 

However, Eq.(\ref{coef}) can still be a reasonable
approximation to obtain the state with the lowest energy in the $d$-band.
This state necessarily has small coefficients for the exceptional registers
$\phi (n_0-1, n_0)$ and $\phi(n_0 +1, n_0+2)$, so the effects caused by their
different energies, which move the eigenstates away from 
the ``Bell-type'' states, are not so significant. Such ``ground'' state
of the band has, as expected and also confirmed numerically, the largest 
concurrence. It can be found analytically
by using Eqs.(\ref{coef}),(\ref{bound_a}), 
and some new boundary conditions for the chain of ordinary qubits.
This is a relatively good approximation to the actual state and provides
a good description for the decay of $C_{\max}$ with $\Delta $.

Since we are interested in the $d$-band ground state, we select $k_1=2$, 
which gives the smallest $\cos\alpha$. The boundary conditions
for the coefficients $b(m)$ are now different.
When the first excitation is on site $n=n_0$, 
the second excitation,
necessarily out of the defects, has to satisfy

\begin{equation}
b(n_0+1)=0.
\end{equation}
Moreover, in this case, the second excitation
is also subjected to the condition

\begin{equation}
J\Delta b(n_0-1+L)+ \frac{J}{2}[b(n_0-2+L)] = E_b b(n_0-1+L),
\end{equation}
which, when combined with the general equation,

\begin{equation}
\frac{J}{2}[b(m-1) + b(m+1)] = E_b b(m),
\end{equation}
leads to

\begin{equation}
J\Delta b(n_0-1+L) =\frac{J}{2} b(n_0+L).
\end{equation}

Equivalently, when the first 
excitation is on site $n=n_0+1$, the boundary conditions
for the second one are

\begin{eqnarray}
&&J\Delta b(n_0+2) =\frac{J}{2} b(n_0+1), \nonumber \\
&&b(n_0+L)=0.
\end{eqnarray}

The wave function in the case of $\Delta <d/J$ is then written as

\begin{eqnarray}
|\psi(n,m)\rangle & = & A \sum_{m=n_0+2}^{n_0-1+L} [S(n_0+L-m)\phi(n_0,m) \nonumber \\
& & +S(m-n_0-1) \phi(n_0+1,m)],
\label{psi}
\end{eqnarray}
where A is a normalization constant and the
function $S(x)$ is given by

\begin{equation}
S(x) = 2\Delta \sin [\beta (x-1)] - \sin[\beta x].
\label{S(x)}
\end{equation}
The $L-2$ different and non-trivial values of $\beta $
are obtained from

\begin{equation}
2\Delta \sin [\beta (L-2)] = \sin[\beta (L-1)].
\label{beta}
\end{equation}

We notice once again that we are only interested in the value
of $\beta$ that gives the smallest $\cos \beta$. The eigenvalues obtained with 
Eqs.(\ref{energy}),(\ref{alpha}), and (\ref{beta}) 
are fairly similar to the ones
obtained numerically, but the results for the eigenvectors (\ref{psi}) are 
so not good. The approximation used here is only acceptable for the 
lowest-energy state of the $d$-band.
Its reduced density matrix becomes

\begin{equation}
  \rho =  
    \left(
	\begin{array}{cccc}
            0 & 0 & 0 & 0 \\
	    0 & 1/2 & |A|^2 s & 0 \\
            0 &|A|^2 s & 1/2 & 0 \\
	    0 & 0 & 0 & 0   	
	\end{array}
    \right),
\end{equation}
where

\begin{equation}
s=\sum_{m=n_0+2}^{n_0-1+L} S(n_0+L-m)S(m-n_0-1).
\end{equation}

The maximum concurrence obtained, 

\begin{eqnarray}
C_{\max}&=&\sqrt{|A|^4 s^2 + |A|^2 s +1/4} \nonumber \\
& & - \sqrt{|A|^4 s^2 - |A|^2 s +1/4},
\end{eqnarray}
is clearly smaller than 1. $|A|^2 s$ can only be equal to $1/2$
when $\Delta =0$ and $|S(n_0+L-m)|=|S(m-n_0-1)|$, that is when all
quantum registers forming the $d$-band are in resonance. 
However, $S(n_0+L-m)$ and $S(m-n_0-1)$ differ mostly at the 
borders of the chain of ordinary qubits (where $m=n_0+2$ or $m=n_0+L-1$) 
and this difference increases as $\Delta $ increases.

In the case of 8 sites,
$d/J=10$ and $\Delta =3$, for example, we find 
$C_{\max}\sim 0.91$, while numerically we have $C_{\max}\sim 0.98$. 
For a longer chain with 12 sites and the same parameters, 
$C_{\max}\sim 0.97$,
while numerically we have $C_{\max}\sim 0.99$. 
The agreement between numerics and analytics are not excellent, 
but the approximation gives the correct trend: the concurrence
decreases as $\Delta $ increases (with $\Delta <d/J$), and
the effects become less important in larger chains.

\subsubsection{Two excitations and $\Delta = d/J$}

In the case of resonance, when $\Delta = d/J$, the quantum registers
$\phi (n_0+L-1, n_0)$ and $\phi(n_0 +1, n_0+2)$ 
have very large energy and do not participate 
in the formation of the eigenstates in the defect band anymore. However,
quantum registers corresponding to the bound pairs 
$\phi (n_0-2, n_0-1)$ and $\phi(n_0 +2, n_0+3)$ now
play an important role, for they are in resonance with
the other registers and couple with them in first order 
of perturbation theory in $J$.
As a result, a transition from the sites $(n_0+1, n_0+3)$
to $(n_0+2, n_0+3)$ [equivalently from $(n_0-2, n_0)$ to $(n_0-2, n_0-1)$], 
which {\it removes} the excitation from the defect, can happen.
This is the cause of the drop in the value of the maximum concurrence.

The scattering process is schematically shown below. There are
four states out of the core that could lead to a Bell-type state.

\footnotesize
\[
\begin{array}{ccccc}
& & &n_0\!-\!1,n_0\!+\!L\!-\!2& \\

& & &\updownarrow&\\

n_0,n_0\!+\!2 & \leftrightarrow\! n_0,n_0\!+\!3\!\leftrightarrow& 
... &n_0,n_0\!+\!L\!-\!2 & \\

& \updownarrow & & \updownarrow & \\

& n_0\!+\!1,n_0\!+\!3\!\leftrightarrow& 
\!...\! & \!n_0\!+\!1,n_0\!+\!L\!-\!2\! &\!\!\!\!\!\leftrightarrow\! 
\!n_0\!+\!1,n_0\!+\!L\!-\!\!1\\

& \updownarrow & & &\\

&n_0\!+\!2,n_0\!+\!3& & &

\end{array}
\]
\normalsize

Notice that even though the other bound pairs far from the defects
are also in resonance with the above states, the coupling with them 
occurs in second order in $J$, so they form a
separated band. A more thorough discussion 
can be found in Ref.\cite{prb}, but the idea is the following.
There are two possible channels for the second order coupling between 
distant bound pairs and the $d$-band states. A possibility is a transition
from sites $(n_0+3, n_0+4)$ to $(n_0+1, n_0+4)$, but we can equivalently hop
from $(n_0+3, n_0+4)$ to the bound pair next to the defect $(n_0+2, n_0+3)$
[two channels also exist for the excitations on sites $(n_0-3, n_0-2)$].
The amplitudes for these two transitions are equal in magnitude, but
opposite in sign. Due to such quantum interference, or antiresonance,
we have two uncoupled bands of very close energy.

The antiresonance explains why the drop in the value of the maximum
concurrence when $\Delta = d/J$ is not so drastic 
as one might have expected. Also, since there are only
four states preventing the creation of a perfect Bell state, the decay of
$C_{\max}$ becomes less perceptible in larger chains.

\subsubsection{Two excitations and $\Delta \gg d/J$}

When $\Delta \gg d/J$, the allowed transitions, shown below, 
now have two registers
out of the core that could lead to a Bell-type state.

\footnotesize
\[
\begin{array}{ccccc}
n_0,n_0\!+\!2& \leftrightarrow n_0,n_0\!+\!3 \leftrightarrow  
& ... & \!n_0,n_0\!+\!L\!\!-\!2 & \\

 & \updownarrow & & \updownarrow  & \\

 & n_0\!+\!1,n_0\!+\!3 \leftrightarrow  
& ... & \!n_0\!+\!1,n_0\!+\!L\!-\!2 \!\! \leftrightarrow \!\!\!\!\!& n_0\!+\!1,n_0\!+\!L\!\!-\!1

\end{array}
\]
\normalsize
This justifies why the maximum concurrence never goes back to 1, even
for very large $\Delta$. However, the effects caused 
by just these two isolated registers, 
$\phi (n_0,n_0+2)$ and $\phi(n_0+1,n_0+L-1)$,
become less significant in larger chains.

\subsubsection{More than two excitations and $\Delta \gg d/J$}

When more than two excitations are present, the analysis of the 
maximum concurrence for small $\Delta$'s gets more complicated. A more
interesting and general situation emerges when $\Delta \gg d/J$. Here,
we have again Bell-type states similar to the ones given by Eq.(\ref{bell}).
These states have one excitation hopping between the two defects and all the
others bound together in a cluster in sites far from the defects. 

Only clusters with the same number of excitations are coupled and
the transitions between them happen in a high order of 
perturbation theory in $J$. 
These transitions involve virtual steps where a dissociation occurs. 
Therefore the clusters move together as a whole and
very slowly. 
We have found that states involving clusters at least two sites away
from the defects have the largest concurrences. They
are written as

\begin{widetext}
\begin{eqnarray}
& & |\psi(n,m_1,m_2,...,m_{N-1})\rangle  =  
\sum_{m_1=n_0+4}^{n_0+L-N-1} 
A \sin \left[ \frac{\pi k_2 (m_1 - n_0 -3)}{(L-N-3)}\right] \times \nonumber \\
& & \frac{1}{\sqrt{2}}\left[ \phi(n_0,m_1,m_1+1,..., m_1+N-2) 
+(-1)^{k_1 +1} \phi(n_0 +1,m_1,m_1+1,...,m_1+N-2) \right],
\label{bell2}
\end{eqnarray}
\end{widetext}
where $k_1=1,2$ and $k_2=1,2,..., L-N-4$. 

Notice that the states above cannot be obtained when we have
small chains with $L\leq 8,9$ and the total number of excitations are
$N=L/2,L/2+1$ respectively, as the available $L-6$ sites are not
enough for the remaining $N-1$ excitations. But, 
for these small chains and also for larger ones, there are other sorts of 
Bell-type states that can be obtained making use of the anisotropy.
An example is the state $1/\sqrt{2}[
\phi(n_0,n_0+2,n_0+3,n_0+L-2,n_0+L-1) + 
\phi(n_0+1,n_0+2,n_0+3,n_0+L-2,n_0+L-1)]$
found when $L=10$. We also note that, 
in the case of $L=12$ and $N=5$, the states with 
large concurrences obtained numerically did not correspond to what was expected
from (\ref{bell2}), though they were also Bell states.

As a result, the anisotropy coupling may be harmful when
small, but very useful when sufficiently large, for we can 
recover Bell-type states.

\begin{figure}[!ht]
\includegraphics[width=2.5in]{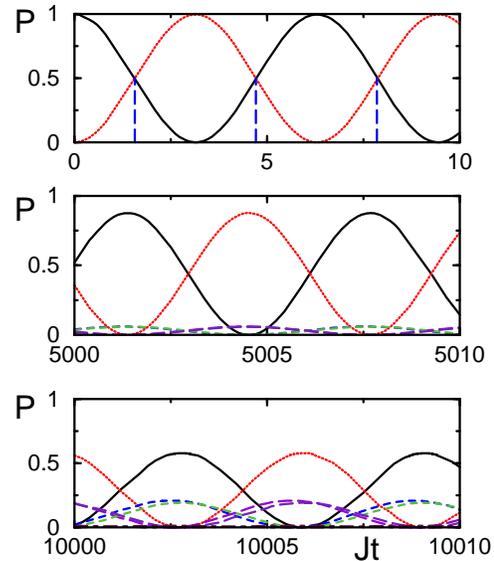}
\caption{(color online) We have a chain with 12 sites, 4 excitations,
and two defects placed on sites 1 and 2.
The parameters of the system are $d=10 J$ and $\Delta = 50$. We 
choose $\phi (1,6,7,8)$ as the initial state and let the system evolve.
The three panels show the probability for finding other 
quantum registers in time. For the short times of the top panel,
$\phi (1,6,7,8)$ (black solid line) mixes only with $\phi (2,6,7,8)$
(red dotted line). The instants of time where Bell-type states 
are created are shown with vertical long dashed lines. They appear
at $\pi/2J$,  $3\pi/2J$, and  $5\pi/2J$. In the middle panel, maximally 
entangled states are not obtained anymore, as other quantum registers
start mixing with the previous two. The probabilities for the
registers  $\phi (1,5,6,7)$ and $\phi (1,7,8,9)$ almost coincide and
are shown with dashed lines (blue and green, respectively). The
same happens to the registers $\phi (2,5,6,7)$ and $\phi (2,7,8,9)$,
which are indicated with long-dashed lines (violet and indigo, respectively).
The lower panel is obtained for times even longer, where the
quantum registers $\phi (1,8,9,10)$ and $\phi (2,8,9,10)$ 
barely start appearing. They are shown in the figure with
dot-dashed (maroon) lines, but can hardly been seen.}
\label{fig2}
\end{figure}

In the limit of very large $\Delta $ we can use the
formation of clusters to create maximally entangled states that 
will remain as such for a long time. As an illustration, we take the 
case of $L=12$, $4$ excitations, $d/J=10$, $\Delta =50$ and the two defects
placed on sites $1$ and $2$.  
The most
straightforward  method to study the dynamics of the system consists
of diagonalizing the $495\times495$ Hamiltonian (\ref{XXZ}) 
for the chosen parameters.
We then assume that the initial state is the quantum 
register $\phi (1,6,7,8)$ 
with one excitation on site $1$ and the other three excitations
bounded in a cluster on sites $6, 7$, and $8$, 
which is not an eigenstate of the Hamiltonian $H$. 
Its evolution is obtained by writing it as a linear superposition of the 
eigenstates of $H$ with their respective energies. The results are shown
in Fig.~\ref{fig2}.

The excitation on site $1$ hops between the two defects
with a period $2\pi /J$ \cite{pra}. The analytical reason for this is the following.
The states with one excitation on one of the defects and three excitations bounded
together form their own energy band around the value 
$\varepsilon_4 +d+J\Delta$.
These states are not coupled with the other states
of the system and can be treated separately, 
largely reducing the Hamiltonian to be considered.
In first 
order of perturbation theory, the Hamiltonian is still further reduced to a simple
$2\times2$ matrix written in the basis of the two registers 
$\phi(1,6,7,8)$ and $\phi(2,6,7,8)$, which are coupled in first order. The off diagonal
elements of this matrix are $J/2$. Trivially, 
the eigenstates and eigenvalues are the EPR
states $\psi_{\pm} =1/\sqrt{2}[\phi (1,6,7,8) 
\pm \phi (2,6,7,8)]$ and $E_{\pm}=
\epsilon_4 +d+J\Delta \pm J/2$, respectivelly. For the initial state considered,
the probability to find the excitation on site $1$ at time $t$ is given by
$P_{\phi(1,6,7,8)}(t)=\{ 1+\cos [(E_+ - E_-)t] \}/2$, 
while the probability to find it in $2$ is
$P_{\phi(2,6,7,8)}(t)=\{ 1-\cos [(E_+ - E_-)t]\} /2$.

At each instant of time
$k\pi /2J$, where $k$ is an odd number, a state of maximal entanglement
between the defects is obtained, $1/\sqrt{2}[\phi (1,6,7,8) 
\pm \phi (2,6,7,8)] $. 
After a very long time, the states $\phi (1,6,7,8)$ and 
$\phi (2,6,7,8)$ finally
start mixing with the other six registers (see Eq. (\ref{bell2})), 
where the cluster appears in a different position.
The hybridization of these states in time is shown in Fig.~\ref{fig2}. 
The top panel corresponds to a short time
and the two lower panels are obtained after much longer times
have passed.

Before allowing the system to evolve for a very long time, where the
cluster delocalizes, states with concurrence 
very close to 1 can be created.
We let the system evolve just up to one of the initial instants where
a Bell state emerges. At this moment, the two defects can
be quickly detuned and such maximally entangled state 
between the defects would be maintained for
a long time until the cluster starts moving.

It is clear that if the initial state was one of 
the EPR states $1/\sqrt{2}[\phi (1,6,7,8) 
\pm \phi (2,6,7,8)] $, 
it would take a very long time to change, since this 
would require moving the ``heavy" cluster of three excitations
together. The ability to keep the entanglement 
between the defects very large
for a long time is a consequence of the large
anisotropy of the $XXZ$ model. In this way, this result 
strongly differs from the dynamics considered in
Ref.~\cite{Amico}. There, such effects of 
the Ising interaction are not seen, because 
the dynamics of correlations is studied for an $XY$ model.

\subsection{Next nearest neighbor defects}

The description of the dependence of the maximum concurrence on the 
anisotropy coupling when the defects are next nearest neighbors
is very complex, especially when
several excitations are present. Just like in the end of the previous 
subsection, we restrict our study to the case of $\Delta \gg d/J$, where
general remarks can be made. 

The states with the largest concurrences are similar to the states 
given by Eq.(\ref{bell2}).
They have one excitation hopping between the
two defects, though they are now separated, so the transition occurs 
in second order in $J$. 
The other excitations are bound together in clusters 
placed at least two sites
away from the defects.
Compared to the case of nearest neighbor defects,
these remaining $N-1$ excitations have now one site less available, 
there are $L-7$ sites for them.

\begin{figure}[!ht]
\includegraphics[width=2.5in]{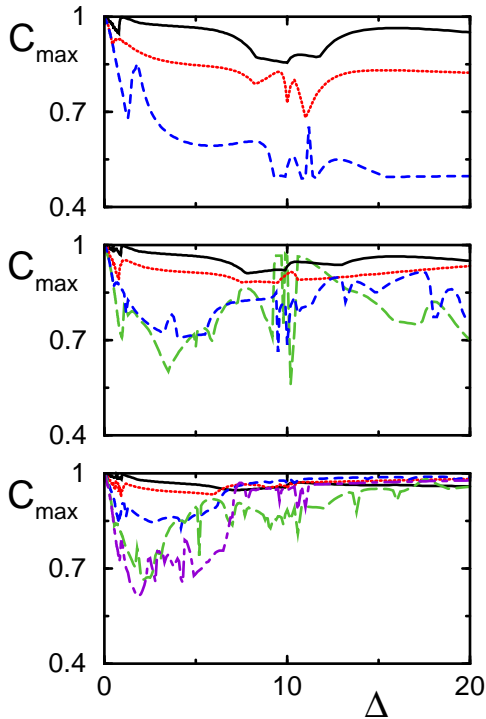}
\caption{(color online)
Maximum concurrence vs. anisotropy coupling in closed Heisenberg spin chains
with defects located on sites 1 and 3. The energy difference between
the defects and the other qubits is $d=10J$. 
The panels correspond to data for 
chains of different lengths: $L=8$ (top), $L=10$ (middle) and 
$L=12$ (bottom). The maximum number of excitations considered for each
chain is $L/2$. The curves are chosen as follows:
solid (black) curves for 2 excitations,
dotted (red) for 3 excitations, dashed (blue) for 4 excitations,
long dashed (green) for 5 excitations, and dot-dashed (violet) 
for 6 excitations.}
\label{fig3}
\end{figure}

When $N-1$ is larger than $L-7$, the states
with the largest concurrence have a different form. Sometimes these
states can also have large concurrences, but this is not always the case, 
which explains the curves 
with low concurrence in the top and middle panels in Fig.~\ref{fig3}.

The important result of this section is the verification that
Bell-type states can again be obtained in the limit of large
chains and $\Delta\gg d/J\gg 1$, even though the defects are 
separated.

An analysis for the time evolution of the system with a particular
$d$-band quantum register taken as the initial state could again be made. It
would be very similar to the one from the previous subsection, though
a longer time is required for a maximally entangled state to appear.

\section{Conclusion}

We have shown how the interplay between interaction and disorder may
affect the entanglement between two defects in a chain of
qubits with anisotropic coupling. When the difference in energy between
the defects and the ordinary qubits $d$ is much larger than the 
hopping integral $J$, several situations may be identified according to the
value of the anisotropy coupling $\Delta $, 
the number of excitations in the chain, where the defects are placed, and 
the chain size. 

The most general case refers to large chains and $\Delta \gg d/J \gg 1$,
where Bell-type states are obtained even for defects that are not 
nearest neighbors. These states have one excitation hopping between the 
two defects and all the others bound together in a cluster far from the 
defect sites. These clusters are a consequence of the large anisotropy. They 
move together and very slowly, allowing the maintenance of a large
entanglement between the defects for a long time.

In the case of nearest neighbor defects, we also verify that even when
the anisotropy coupling is not very large, the decrease in the 
value of entanglement between the 
defects is never very abrupt and it becomes less significant in larger
chains. Analytical results were obtained for 
two excitations when $\Delta=0$ and $\Delta <d/J$. 
For $\Delta <d/J$ the state with the largest concurrence is the one
with the smallest energy in its band.

We note that in this paper we have not developed any analysis in terms
of quantum phase transition as done in several previous works \cite{osterloch,osenda,huang}. By selecting 
two very large defects we guarantee that
the entanglement between them is mostly kept very large. If only the $XX$ part
of the Hamiltonian (\ref{XXZ}) was present, i. e. if $\Delta = 0$, their concurrence 
would always have the maximum value 1, as seen from 
Figs.~\ref{fig1} and \ref{fig3}. It is
the effect of the extra Ising interaction that can sometimes decrease
the entanglement. 

The present work has its clear relevance for quantum information and quantum 
computing, but it should also be of interest for condensed matter 
physics, where one wants to understand how localization may be affected
by interaction and disorder.

\acknowledgments 
L. F. S. acknowledges support by the NSF through grant No. ITR-0085922 and
thanks M. I. Dykman for discussions. G. R. thanks
FAPESP for funding this research. We are both very grateful to C. O. Escobar
for helpful suggestions.

\end{document}